\documentclass[%
 reprint,
 amsmath,amssymb,
 aps,
]{revtex4-1}

\usepackage[final]{pdfpages}

\usepackage{graphicx}
\usepackage{dcolumn}
\usepackage{bm}
\mathchardef\mhyphen="2D

\makeatletter
\AtBeginDocument{\let\LS@rot\@undefined}
\makeatother

\begin{document}

\preprint{APS/123-QED}

\title{Integrated Kerr frequency comb-driven silicon photonic transmitter}

\author{Anthony Rizzo\textsuperscript{1,$\dagger$}, Asher Novick\textsuperscript{1,$\dagger$}, Vignesh Gopal\textsuperscript{1,$\dagger$}, Bok Young Kim\textsuperscript{2}, Xingchen Ji\textsuperscript{1}, Stuart Daudlin\textsuperscript{1}, Yoshitomo Okawachi\textsuperscript{2},  Qixiang Cheng\textsuperscript{1,3}, Michal Lipson\textsuperscript{1}, Alexander L. Gaeta\textsuperscript{1,2}, and Keren Bergman\textsuperscript{1,*}}%

\address{\textsuperscript{1}Department of Electrical Engineering, Columbia University, New York, NY 10027 \\
\textsuperscript{2}Department of Applied Physics and Applied Mathematics, Columbia University, New York, NY 10027 \\
\textsuperscript{3}Current Address: Engineering Department, University of Cambridge, Cambridge CB3 0FA, UK \\
\textsuperscript{$\dagger$}These authors contributed equally to this work }

\date{\today}

\begin{abstract}
The exponential growth of computing needs for artificial intelligence and machine learning has had a dramatic impact on data centre energy consumption, which has risen to environmentally significant levels. Using light to send information between compute nodes can dramatically decrease this energy consumption while simultaneously increasing bandwidth. Through wavelength-division multiplexing with chip-based microresonator Kerr frequency combs, independent information channels can be encoded onto many distinct colours of light in the same optical fibre for massively parallel data transmission with low energy. While previous demonstrations have relied on benchtop equipment for filtering and modulating Kerr comb wavelength channels, data centre interconnects require a compact on-chip form factor for these operations. Here, we demonstrate the first integrated silicon photonic transmitter using a Kerr comb source. The demonstrated architecture is scalable to hundreds of wavelength channels, enabling a fundamentally new class of massively parallel terabit-scale optical interconnects for future green hyperscale data centres.
\end{abstract}

\maketitle

With the rise of cloud-based computing, computational workloads have largely been offloaded from local machines onto the server racks of hyperscale data centres and high-performance computers. Bandwidth-hungry applications such as artificial intelligence and machine learning have severely strained the interconnects within these systems, creating a threat of scaling stagnation without significant changes to the physical hardware used to pass data between nodes \cite{Cheng:18, Miller2000}. Furthermore, the energy consumption of such data centres has become environmentally significant \cite{strubell2019energy} and will be dominated by interconnect energy in the aforementioned communication-intensive workloads. Optical interconnects based on silicon photonics have been widely recognized as a promising avenue for interconnects to keep pace with these ever-growing bandwidth demands while additionally decreasing energy consumption compared to their electrical counterparts. This promise is largely due to the compact footprint of silicon-on-insulator (SOI) devices, the compatibility of SOI photonics processes with the ubiquitous complementary metal-oxide-semiconductor (CMOS) infrastructure used to fabricate electronic chips, and the tremendous inherent parallelism provided by optics through dense wavelength division multiplexing (DWDM) \cite{Miller2009, miller_attojoule, Sun2015, Atabaki2018}. In particular, DWDM enables independent information channels to be encoded on different colours of light simultaneously in a single optical waveguide or fibre, in stark contrast to copper wire-based links which only permit a single channel per physical connection.

The use of frequency combs for DWDM optical interconnects is an appealing prospect due to their ability to replace the currently-used large arrays of lasers with a single multi-wavelength source \cite{Hu2021, Gaeta2019}. For massively parallel wavelength scaling, microresonator-based Kerr frequency combs in the silicon nitride (Si\textsubscript{3}N\textsubscript{4}) platform show particular promise due to their compact size, CMOS compatibility, and ability to generate hundreds of evenly-spaced low-noise wavelength channels from a single continuous-wave (CW) laser source \cite{Levy2010, Levy2012, Kippenberg18}. Chip-based frequency combs have been widely demonstrated for long-reach optical communications using benchtop telecommunications equipment for filtering, modulating, and receiving wavelength channels \cite{Marin-Palomo2017, Pfeifle2014, Fulop2018, Kong:20, Hu2018, Corcoran2020}. However, DWDM has drawn a great deal of interest for data centre interconnects \cite{Ayar2020, Liang:20, Chen:15}, which place stringent requirements on the energy consumption and footprint of transceivers. 

Notable demonstrations with semiconductor mode-locked lasers \cite{Chen:15, soa2017} and distributed feedback laser (DFB) arrays \cite{Zheng:13} have shown high bandwidth data communication with compact silicon photonic transmitter chips, but used modest wavelength channel counts (five, eight, and eight, respectively) and lack the inherent scalability provided by Kerr combs. In particular, semiconductor mode-locked lasers are fundamentally restricted in wavelength scaling by the gain bandwidth of the active region material \cite{Liu:19}, while DFB arrays suffer from non-uniform channel spacing due to fabrication variations between the discrete laser cavities as well as the need to multiplex all of the channels onto a single output fibre. A monolithic lithium niobate chip was used to demonstrate Kerr comb generation, filtering, and modulation on a single die, but was restricted to a single filtered channel at megabit/s data rates due to the noisy modulational instability state of the comb \cite{Wang2019}. 

Here, we demonstrate the first integrated Kerr comb-driven silicon photonic transmitter using a novel, massively scalable link architecture. The transmitter was designed for 32 wavelength channels and showed open eye diagrams for all channels up to 16 Gb/s, yielding an aggregate single-fibre bandwidth of 512 Gb/s. Furthermore, the generated comb lines showed a negligible power penalty compared to a tunable CW laser source at the same wavelength, demonstrating that each tone behaves identically to an independent CW carrier from an array of lasers. This work represents a novel and realistic direction for data centre interconnects to scale to hundreds of wavelength channels, enabling future multi-terabit/s chip-to-chip links operating at energies below one picojoule per bit.

\begin{figure}
    \centering
    \includegraphics[scale=0.97]{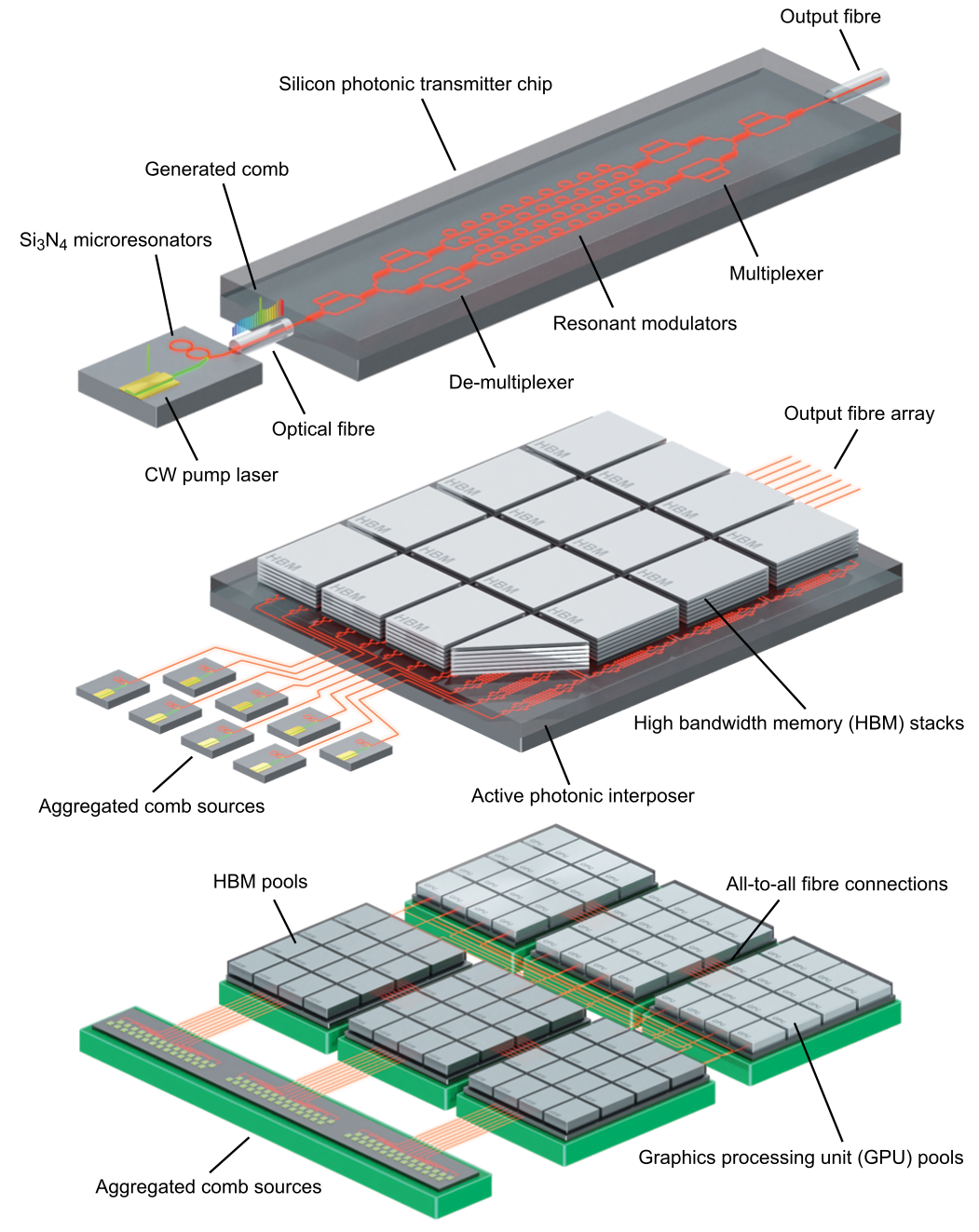}
    \caption{\textbf{Artistic vision of a disaggregated data centre based on Kerr frequency comb-driven silicon photonic links.} Hierarchical view of low-energy terabit-scale links connecting high bandwidth memory stacks to graphics processing units in a future data centre system. The large number of wavelengths provided by integrated microresonator-based Kerr combs enable massively parallel data transmission on a single optical fibre through wavelength division multiplexing. Resources are packaged on active photonic interposers and connected via optical fibre, exploiting the distance-agnostic nature of optical communications to pool and connect spatially-distanced resources with comparable latency and energy consumption to in-package electrical signaling. CW: continuous-wave, Si\textsubscript{3}N\textsubscript{4}: silicon nitride, HBM: high bandwidth memory, GPU: graphics processing unit.}
    \label{fig:artist}
\end{figure}


\begin{figure*}
    \centering
    \includegraphics[scale=0.98]{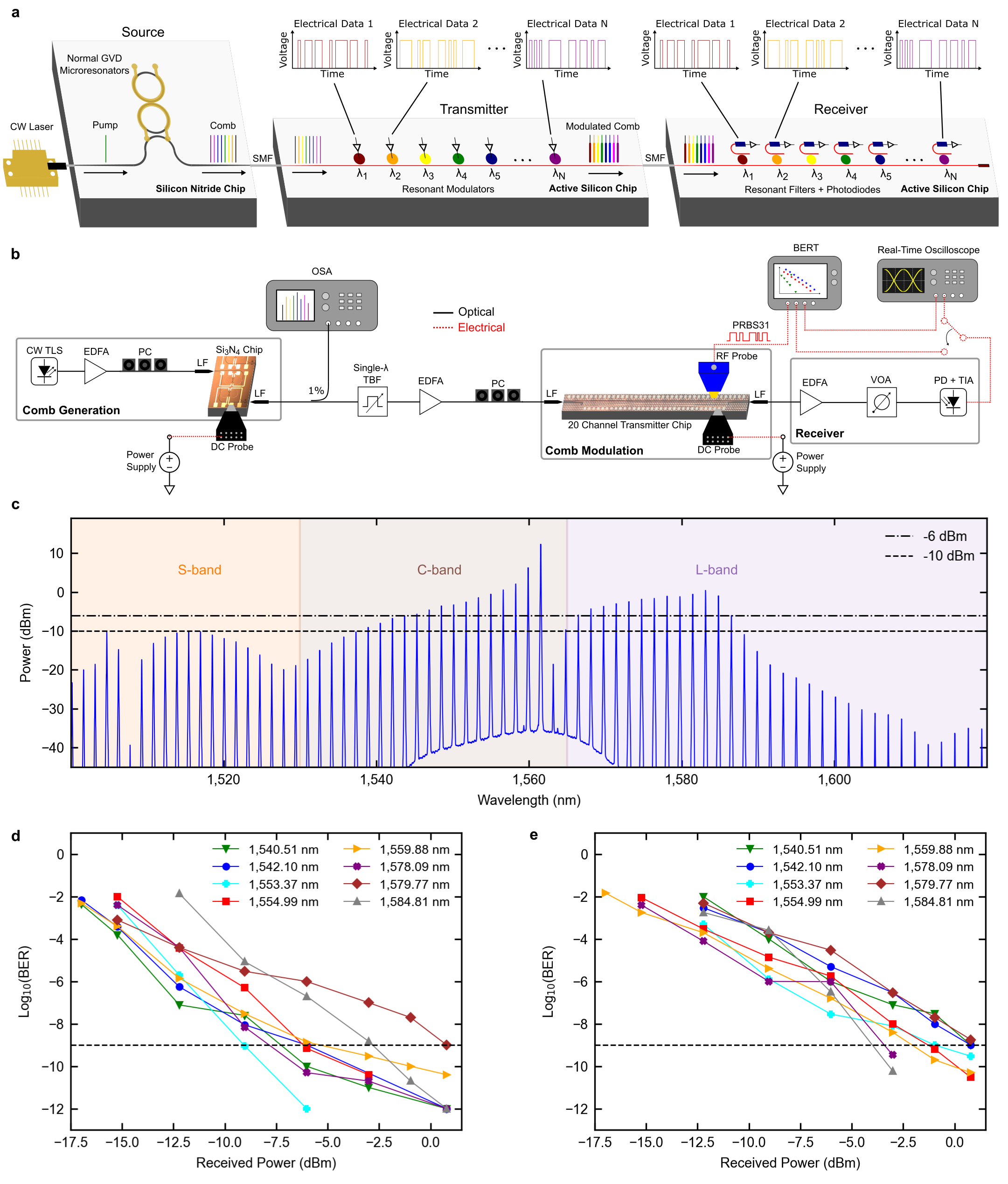}
    \caption{\textbf{Single bus architecture and results.} \textbf{a}, Schematic of a Kerr frequency comb-driven single bus cascaded resonator-based link. CW: continuous-wave, GVD: group velocity dispersion, SMF: single mode fibre. \textbf{b}, Experimental setup used for the single bus data transmission demonstrations. CW TLS: continuous-wave tunable laser source, EDFA: erbium-doped fibre amplifier, PC: polarisation controller, LF: lensed fibre, Si\textsubscript{3}N\textsubscript{4}: silicon nitride, DC: direct current, OSA: optical spectrum analyser, RF: radio frequency, TBF: tunable bandpass filter, VOA: variable optical attenuator, PD: photodetector, TIA: trans-impedance amplifier, BERT: bit-error rate tester, PRBS: pseudo-random bit sequence. \textbf{c}, Measured normal-GVD comb spectrum with markers denoting -6 dBm and -10 dBm thresholds. \textbf{d}, Bit error rate (BER) waterfall curves for eight measured comb lines at 10 Gb/s and \textbf{e}, 16 Gb/s. All lines achieve directly-measured BER better than $10^{-9}$ for both data rates without forward error correction.}
    \label{fig:single_bus}
\end{figure*}

\section*{Results}


While past demonstrations have focused on coherent communications using Kerr microcombs as DWDM sources, high performance computer and data centre interconnects require low latency and low energy per bit, making intensity-modulated direct-detection (IM-DD) solutions more appealing than coherent solutions due to the absence of energy-intensive and latency-inducing digital signal processing (DSP). In particular, IM-DD links with native error-free signaling do not require forward error correction (FEC) and thus have no encoding overhead, easing requirements on the electronics to reconstruct the original bit stream. Additionally, coherent communication links require a local oscillator at the receiver, necessitating another light source which consumes additional energy. 

To emphasize these realistic conditions for future short-reach links, we focus our analysis and experiments on modest single channel data rates (10 -- 16 Gb/s/$\lambda$) using a standard non-return-to-zero on-off-keying (NRZ-OOK) modulation format. Our link analysis indicates that these modest per-channel data rates with massively parallel scaling in the number of wavelength channels per fibre lead to higher energy efficiencies while still maintaining terabit/s per fibre aggregate bandwidths (Supplementary Notes 1 \& 2). Interconnects with high bandwidth, low energy consumption, and low latency are critical for future disaggregated data centre architectures, which are uniquely enabled by the combination of these three properties (Fig. \ref{fig:artist}).

Standard single bus cascaded microresonator-based links enable DWDM by assigning each resonant modulator/filter to a particular wavelength and have sparked widespread academic and commercial interest \cite{Xu:06, London2019, Ayar2020}. The resonance condition of each resonator depends on the effective refractive index of the optical mode and the round-trip path length and is given by \cite{ringresonators}:
\begin{equation}
\lambda_{res} = \frac{n_{eff}L}{m}
\end{equation}
where $\lambda_{res}$ is the resonance wavelength, $n_{eff}$ is the effective refractive index of the fundamental transverse electric (TE) mode, $L$ is the round-trip path length of the resonant cavity, and $m$ is the integer mode number. By cascading resonators of slightly different radii in even steps along a single bus waveguide, different resonances are used to selectively encode data onto different wavelengths with uniform channel spacing. Since the architecture relies on a comb of wavelengths multiplexed on a single bus waveguide for both the transmitter and receiver, this architecture is a natural solution for Kerr frequency comb-driven links \cite{Novick2021} (Fig. \ref{fig:single_bus}a).

\begin{figure*}
    \centering
    \includegraphics[scale=0.98]{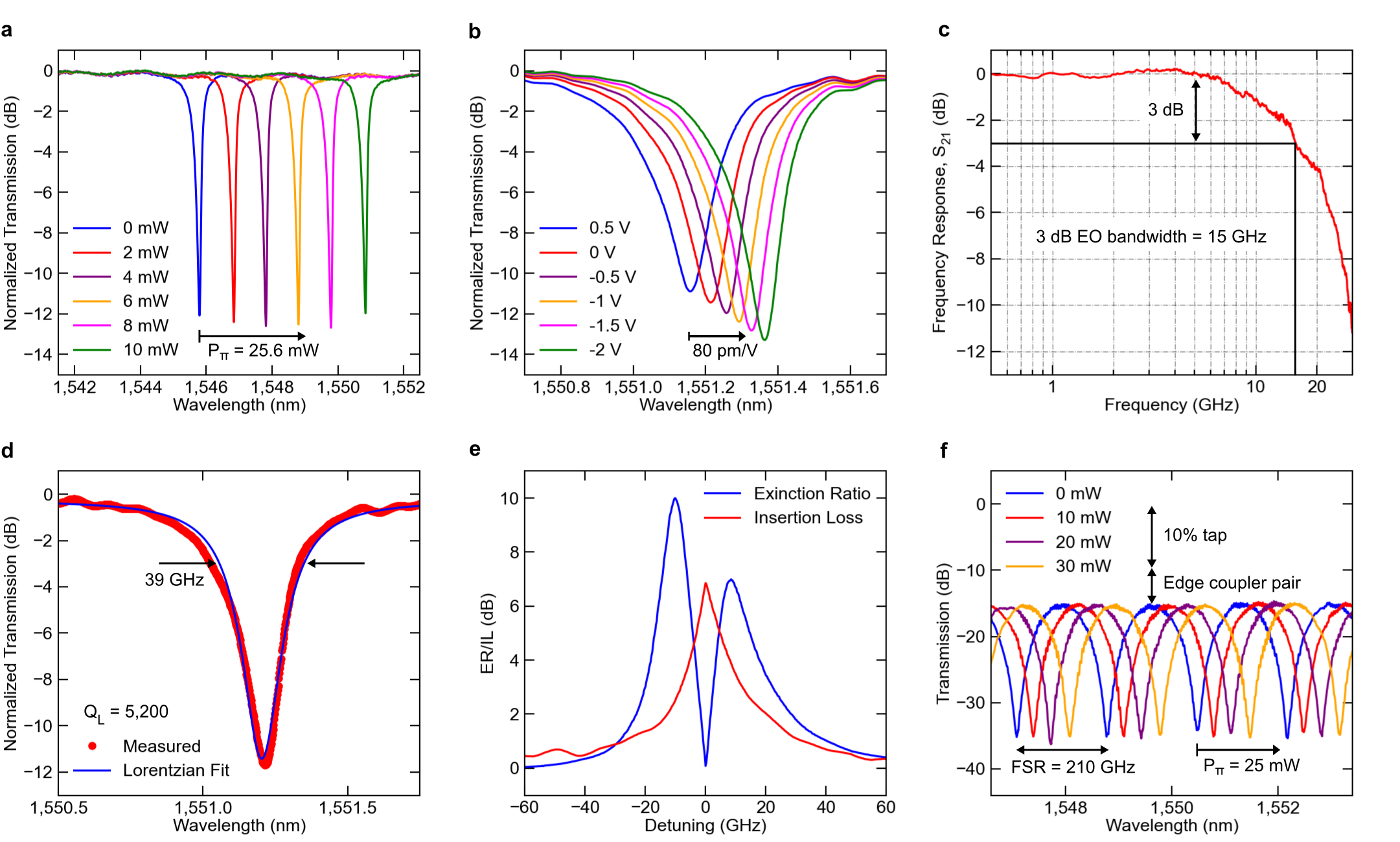}
    \caption{\textbf{Experimental characterisation of photonic devices.} \textbf{a}, Thermo-optic response of the microdisk heaters as a function of dissipated power showing a tuning efficiency of 0.5 nm/mW ($P_{\pi} = 25.6$ mW). \textbf{b}, Modulator depletion response displaying an ultra-high modulation efficiency of 80 pm/V. \textbf{c}, Measured S\textsubscript{21} electro-optic response of the microdisk modulator with a 0 V DC bias demonstrating a 3 dB bandwidth of 15 GHz. \textbf{d}, Measured resonance of the modulator with a Lorentzian fit indicating a loaded quality factor (Q\textsubscript{L}) of 5,200. \textbf{e}, Extinction ratio (ER) and insertion loss (IL) tradeoff for different biasing points of the modulator with a 1.3 V peak-to-peak swing. The asymmetry is due to the increasing Q\textsubscript{L} with reverse bias, which narrows the resonance linewidth and increases the static extinction ratio as it is driven closer to critical coupling. \textbf{f}, Interleaver response with applied power to the bottom arm phase shifter showing a tuning efficiency of $P_{\pi} = 25$ mW and FSR = 210 GHz. The vertical lines indicate a 10 dB loss from the 10\% tap and a 5 dB loss between the edge coupler pair for an inferred coupling loss of 2.5 dB/facet.}
    \label{fig:device_characterization}
\end{figure*}

Integrated thermo-optic heaters in the microdisk modulators are used to tune each resonant wavelength to align to the target comb line. The modulators have an embedded vertical p-n junction which overlaps with the optical whispering gallery mode. When driven in reverse bias, the width of the depletion region is modulated which changes the free carrier distribution interacting with the optical mode and thus changes the effective refractive index through the plasma-dispersion effect. This change in the effective index corresponds to a shift in the resonant wavelength, which modulates the amplitude of the light to achieve levels of `0' and `1' depending on the applied bias. Due to the large overlap between the depletion region and optical mode in vertical junction devices \cite{Timurdogan2014}, the modulators can be directly driven at CMOS-compatible voltages while maintaining a high extinction ratio and low insertion loss (Fig. \ref{fig:device_characterization}e). Furthermore, silicon photonic modulators based on the plasma-dispersion effect have been experimentally shown to function over hundreds of nanometres \cite{Doerr2016} and thus will not restrict the wavelength scalability of the system. For all data transmission experiments in this work, the modulators were driven at 1.3 V\textsubscript{pp}, yielding an extinction ratio of 7 dB with an insertion loss of 3 dB (Fig. \ref{fig:device_characterization}e).  

The source consists of a Si\textsubscript{3}N\textsubscript{4} Kerr comb operating in the normal group velocity dispersion (GVD) regime which demonstrates higher conversion efficiency, power per line, and spectral flatness compared to soliton Kerr combs in the anomalous GVD regime \cite{Kim:19, jang2020, Fulop2018, Lobanov2015, Liu:14, Xue2015}, making normal-GVD combs better suited for data communication applications. The comb was designed to have 200 GHz spacing to operate in the ideal regime for bandwidth and conversion efficiency while maintaining high power-per-line. Similar devices have shown experimentally measured pump-to-comb conversion efficiencies up to 41\% \cite{Kim:19}. The dual resonators have platinum heaters above the Si\textsubscript{3}N\textsubscript{4} waveguide layer, enabling thermo-optic tuning of an avoided mode crossing to enable arbitrary pump wavelengths for generating  the low-noise comb spectrum \cite{Kim:19, Miller:15}. To achieve the low-noise wavelength channels required for high fidelity data transmission, the resonances are tuned to generate a low-noise comb for all experiments (Supplementary Note 3). The fibre-coupled output spectrum of the comb is shown in Fig. \ref{fig:single_bus}c, demonstrating 25 lines above -6 dBm and 33 lines above -10 dBm. The output edge coupler loss on the Si\textsubscript{3}N\textsubscript{4} chip was measured to be -3 dB, leading to 25 lines with on-chip power above -3 dBm.

To evaluate the performance of the single bus architecture with a microcomb source, we fabricated a transmitter chip with 20 cascaded microdisk modulators through the American Institute for Manufacturing (AIM) multi-project wafer (MPW) service (Methods) (Fig. \ref{fig:single_bus}b). Lensed fibre was used to couple the comb output to the transmitter chip input, which displayed a coupling loss of 5 dB/facet and thus required amplification before and after the transmitter. Multiple comb lines were sampled over 44.3 nm of spectral bandwidth in the C- and L-bands, with all eight lines achieving directly-measured bit error rates (BER) better than $10^{-9}$ for 10 Gb/s/$\lambda$ and 16 Gb/s/$\lambda$ without the use of forward error correction (Fig. \ref{fig:single_bus}d,e). 

Despite supporting modest DWDM channel counts, single bus cascaded resonator links run into fundamental scaling restrictions due to the limited free spectral range (FSR) of the resonators, constraining the usable bandwidth to a single FSR. To pack more channels into a single FSR, the channel spacing must be reduced; however, channel spacings below 100 GHz lead to severe crosstalk penalties \cite{kishore} and thus place an upper bound on the number of allowable channels within a given FSR. To mitigate this restriction, we propose a scalable link architecture which sub-divides the comb to reduce the number of channels per bus and permits channel arrangements which are not restricted to a single resonator FSR \cite{Rizzo21} (Supplementary Note 4).

\begin{figure*}
    \centering
    \includegraphics[scale=0.98]{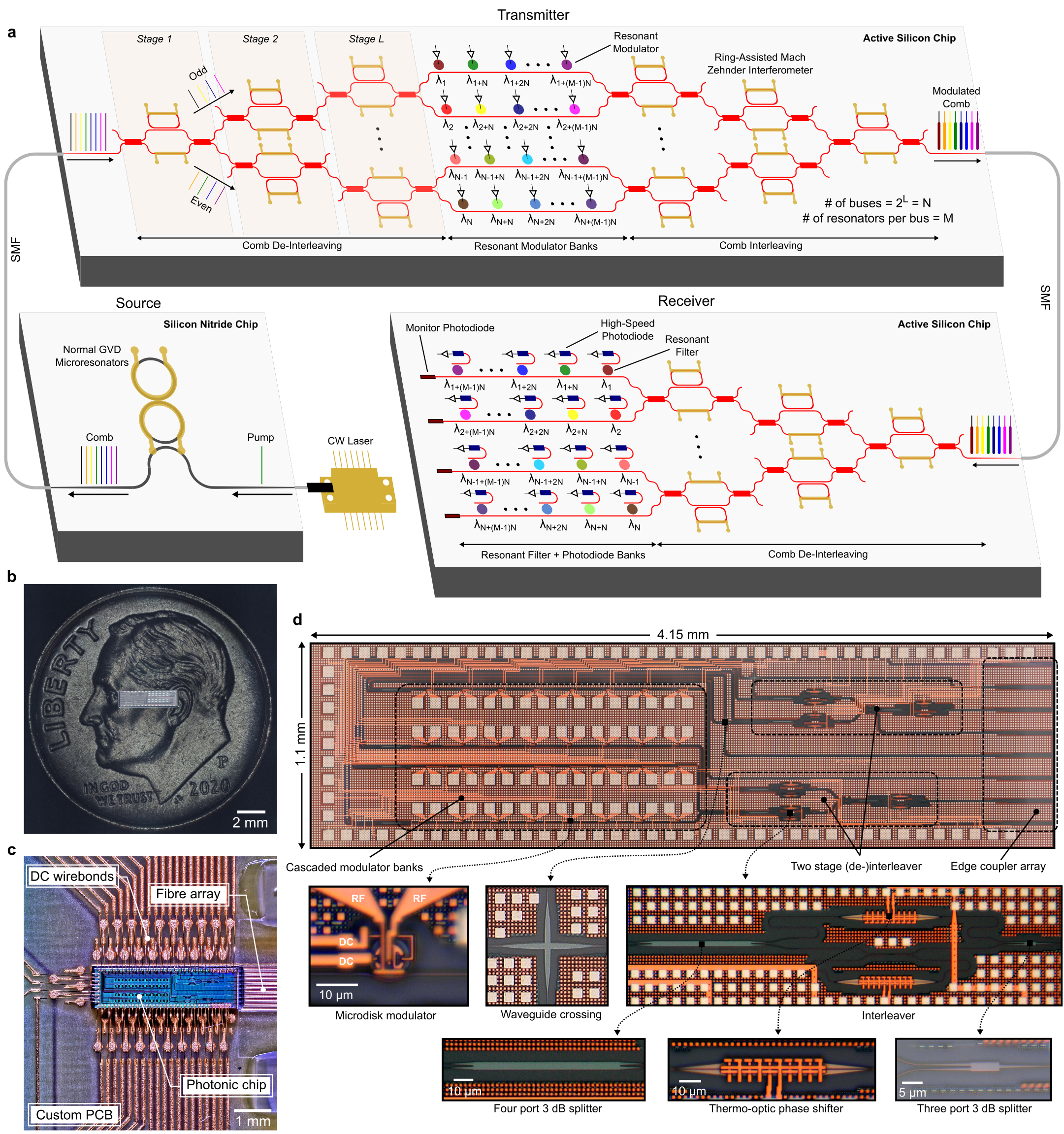}
    \caption{\textbf{Overview of the scalable system architecture and fabricated transmitter chip.} \textbf{a}, Schematic of the proposed scalable link architecture for a highly parallel DWDM silicon photonic interconnect using a broadband Kerr frequency comb source. CW: continuous-wave, GVD: group velocity dispersion, SMF: single mode fibre. \textbf{b}, Micrograph of the unpackaged chip on a U.S. dime. \textbf{c}, Micrograph of the electrically and optically packaged system highlighting key constituent components. DC: direct current, PCB: printed circuit board. \textbf{d}, Micrograph of the full die prior to wire bonding with additional representative micrographs of all active and passive devices on the chip.}
    \label{fig:32channel}
\end{figure*}

The proposed scalable link architecture is shown in Fig. \ref{fig:32channel}a. The source consists of an external or integrated CW laser which pumps a Si\textsubscript{3}N\textsubscript{4} dual-microresonator system to generate the normal-GVD comb source. Standard single mode fibre (SMF) is used to connect the source to the transmitter chip. The transmitter architecture uses a binary tree of asymmetric Mach Zehnder interferometers (MZI) to split the comb spectrum into even and odd groups at each stage. After the comb is divided into subgroups, the groups traverse separate buses of cascaded microdisk modulators, taking advantage of the spectral selectivity of resonant modulators to encode independent data streams onto different coloured comb lines. After each channel is modulated, they are recombined using an identical binary tree of MZIs and coupled off the chip into SMF. At the receiver, the modulated comb lines traverse a final MZI binary tree and are then incident on buses with cascaded resonant filters tuned to match the respective wavelength channels. Each resonant filter has a photodetector at its drop port to convert the optical data streams back into the electrical domain. This architecture naturally enables a disaggregated laser source, which allows the pump and comb to be separately stabilised away from the harsh thermal environment of co-packaged optics and electronics. 

The binary trees of MZI filters serve multiple purposes---depending on the depth of the tree $d$, the channel spacing on each bus is increased by a factor of $2^{d}$. Additionally, the number of channels per bus is reduced by a factor of $2^{d}$, enabling channel configuration schemes that are not restricted by the FSR of the resonant modulators/filters (Supplementary Note 4). Recently, a fully automated control algorithm for aligning similar cascaded MZI binary trees was demonstrated \cite{Akiyama:21}, providing a straightforward path to full-system initialisation and stabilisation. 


To verify the proposed architecture experimentally, we fabricated a 4.15 mm $\times$ 1.1 mm photonic transmitter chip designed for 32 wavelength channels in a commercial 300 mm foundry through the AIM MPW service. The (de-)interleavers were implemented as asymmetric ring-assisted Mach Zehnder interferometers (RMZIs) which exhibit flatter pass- and stop-bands compared to standard asymmetric MZI filters \cite{Luo:10, Rizzo2021}. Two stages of de-interleavers were used to subdivide the comb into four groups which are incident on banks of eight cascaded microdisk modulators (Fig. \ref{fig:32channel}d). After modulation, the groups are recombined onto a single fibre using two stages of RMZI interleavers. While the transmitter chip only requires a single fibre in and a single fibre out for data transmission, six ancillary optical I/O ports are used to tap the circuit at various points for (de-)interleaver and modulator alignment.

\begin{figure*}
    \centering
    \includegraphics[scale=0.98]{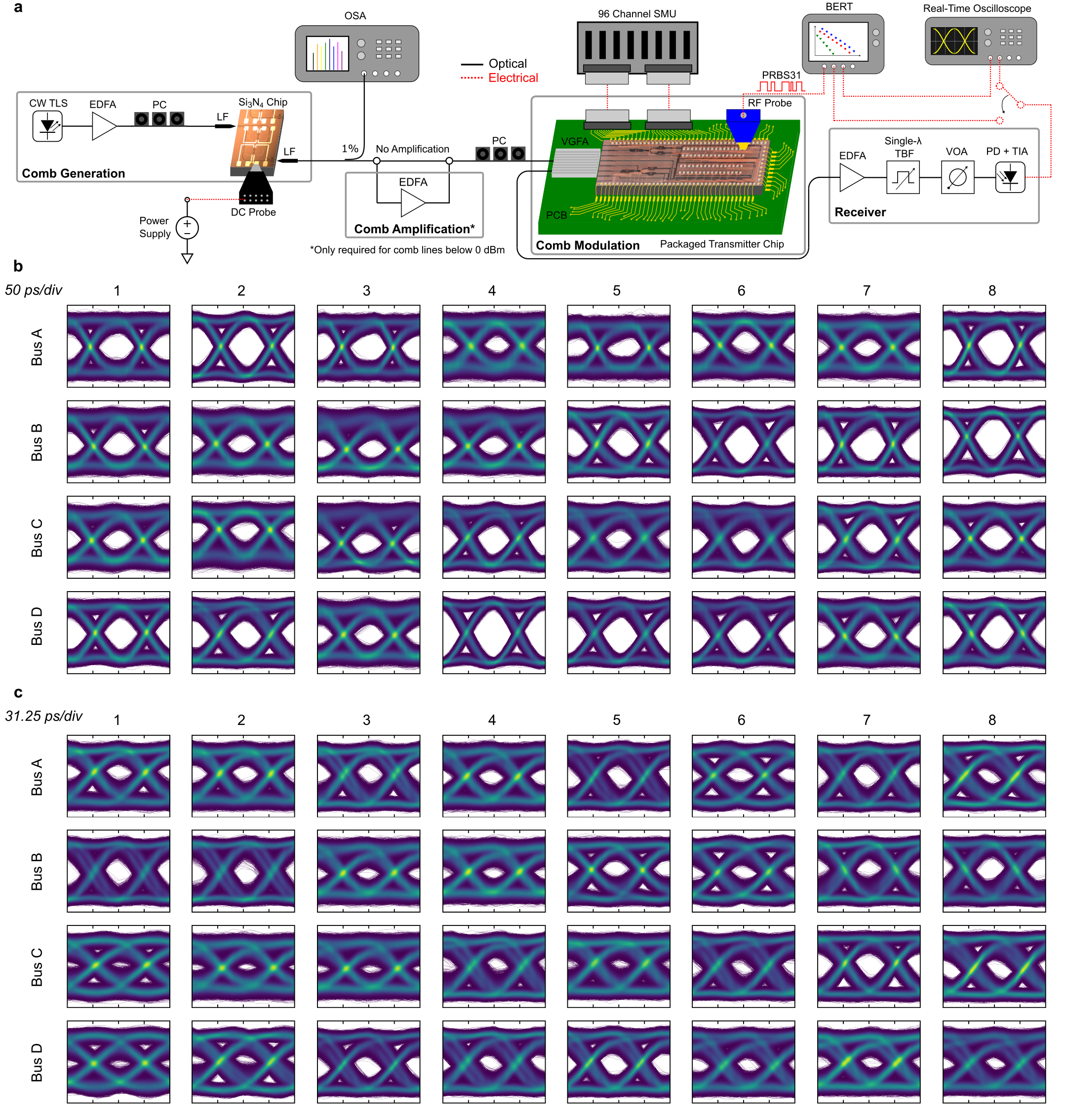}
    \caption{\textbf{32 channel transmitter experimental setup and eye diagrams.} \textbf{a}, Experimental setup used for data transmission experiments. CW TLS: continuous-wave tunable laser source, EDFA: erbium-doped fibre amplifier, PC: polarisation controller, LF: lensed fibre, Si\textsubscript{3}N\textsubscript{4}: silicon nitride, DC: direct current, OSA: optical spectrum analyser, VGFA: v-groove fibre array, SMU: source measurement unit, PCB: printed circuit board, RF: radio frequency, TBF: tunable bandpass filter, VOA: variable optical attenuator, PD: photodetector, TIA: trans-impedance amplifier, BERT: bit-error rate tester, PRBS: pseudo-random bit sequence. \textbf{b}, Eye diagrams for all 32 modulators on the chip at 10 Gb/s and \textbf{c}, 16 Gb/s.}
    \label{fig:experimental_setup}
\end{figure*}

The photonic transmitter chip was optically packaged with an eight channel SMF array at 127 $\mu$m pitch and electrically packaged for the DC thermal controls with 96 wirebonded pads around the periphery at 100 $\mu$m pitch (Methods). The measured optical coupling loss after UV curing with index matching epoxy was 2.5 dB/facet, which greatly reduced the required optical power per line from the comb to close the link budget compared to the bare die case. Due to the high parasitic inductance of wirebonds, the inner high-speed radio frequency (RF) pads to the modulators were left exposed for probing with multi-contact wedge probes. All thermal biases were addressable with uniform current-voltage characteristics, indicating 100\% yield for the DC wirebonds and photonic devices. Furthermore, all modulators on the chip displayed open eye diagrams up to 16 Gb/s (Fig. \ref{fig:experimental_setup}b,c), indicating uniform electro-optic bandwidth across devices (Fig. \ref{fig:device_characterization}c) and 100\% yield for the high speed photonic devices. The perfect yield and high device uniformity emphasise the advantage of fabrication in a commercial 300 mm CMOS foundry and bode well for the system's potential for future high volume scaling.

\begin{figure}
    \centering
    \includegraphics[scale=0.97]{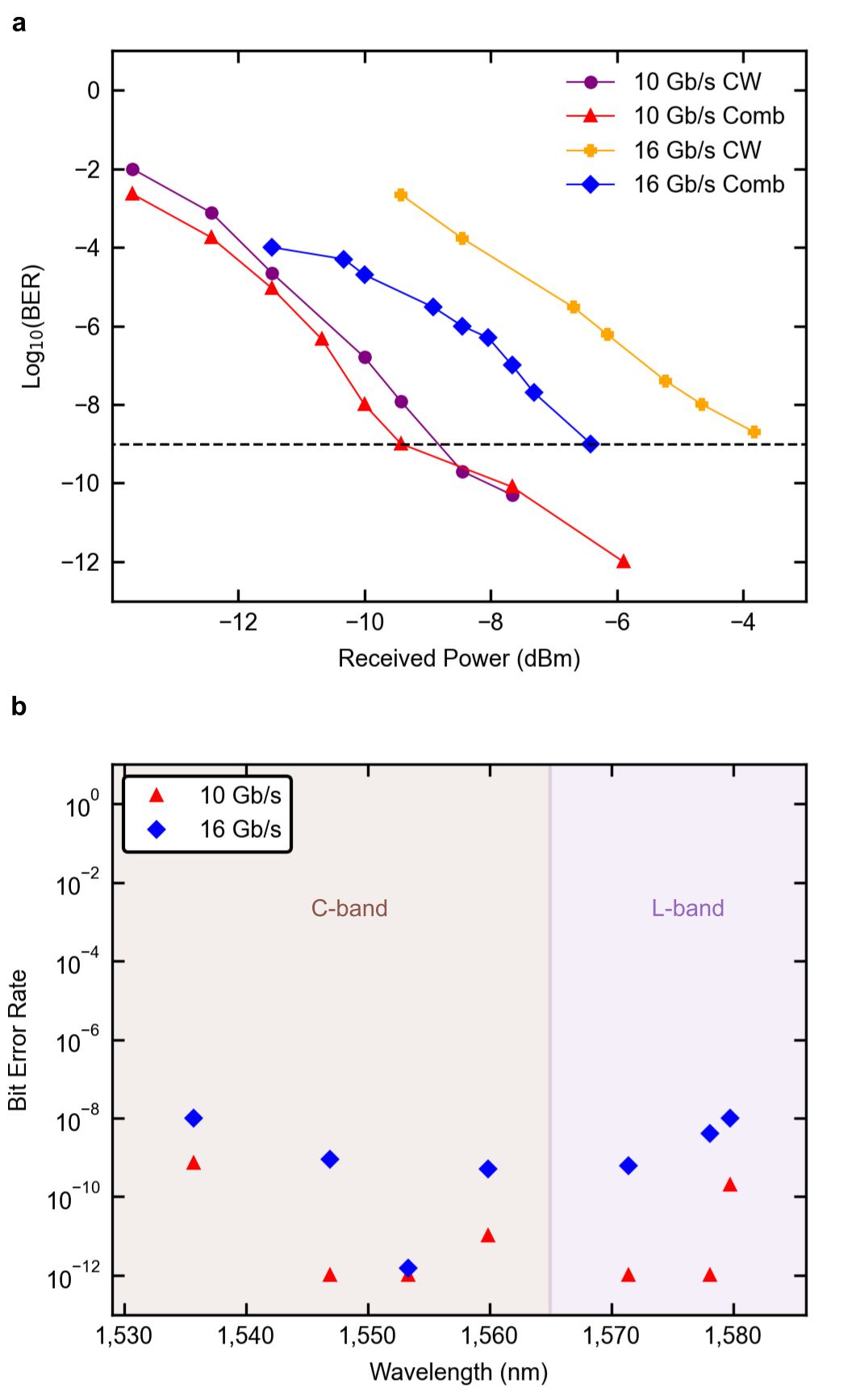}
    \caption{\textbf{32 channel transmitter bit error rate results.} \textbf{a}, Measured BER versus received optical power comparing the performance of a CW laser to a generated comb line at 1,559.8 nm without comb amplification prior to the transmitter chip. \textbf{b}, Measured BER for multiple comb wavelength channels sampled across the C- and L-bands.}
    \label{fig:ber}
\end{figure}

The individual comb lines showed a negligible power penalty compared to a tunable CW laser source and in some cases showed an improvement over the CW source (Fig. \ref{fig:ber}a). Due to the low coupling losses of the packaged transmitter chip and the low on-chip losses of the photonic devices, the comb line one channel blue detuned from the pump (1,559.8 nm) had enough fibre-coupled optical power to close the link budget without comb amplification prior to the transmitter chip (Fig. \ref{fig:ber}a). This result is promising for future link iterations with reduced fiber-chip coupling losses and integrated high sensitivity receiver modules \cite{Daudlin21}, as it shows that the generated comb lines can natively have enough optical power per line to close the link budget without the need for EDFAs (Supplementary Note 1).

To further verify the system performance, BER values were directly measured for sampled comb lines across the C- and L-bands (Fig. \ref{fig:ber}b). Due to the cascaded interleavers on the transmitter chip used for subdividing the comb, no filtering was required between the Si\textsubscript{3}N\textsubscript{4} comb chip and the transmitter package. All off-chip filtering was restricted to the receiver-side, where a tunable filter was used to emulate resonant drop filters for selecting single modulated comb lines.  At both 10 Gb/s and 16 Gb/s, all comb lines achieved BER values better than $10^{-8}$, with most lines below $10^{-9}$ and some natively error-free ($<$ $10^{-12}$). With improved photonic devices, improved receiver sensitivity, and the elimination of noisy EDFAs, it is likely that every comb line across the S-, C-, and L-bands can achieve error-free performance based on their coherence and noise properties \cite{Kim:19}. 


\section*{Discussion}

In conclusion, we have demonstrated the first integrated photonic transmitter using a normal-GVD Kerr comb source operating at a state-of-the-art aggregate bandwidth through massively parallel DWDM. In contrast with other comparable high-bandwidth solutions which scale through increasing the single-channel baud rate and moving to higher-order modulation formats, our demonstrated approach permits modest per-channel data rates with a standard NRZ-OOK modulation format. This approach enables extreme energy efficiency and low latency through elimination of power-hungry DSP and FEC while still maintaining high aggregate bandwidths per fibre. We experimentally characterised data transmission up to 512 Gb/s on a single fibre (465 Gb/mm shoreline bandwidth density) with continued room for scaling through a novel link architecture. Moreover, normal-GVD Kerr combs inherently have high conversion efficiencies ($>$ 30\%) and recently have been shown to synchronize to produce comb line powers otherwise unattainable \cite{kim2021synchronization}.

The transmitter chip was fabricated in a standard SOI process at a commercial foundry, enabling straightforward packaging with modern CMOS electronics for fully integrated transceivers either through heterogeneous \cite{Daudlin21} or monolithic \cite{global} integration. Furthermore, this architecture enables continued scaling in the number of wavelength channels as well as per-channel data rate (with a trade-off in energy efficiency) with improved modulator designs \cite{Sun:19} and is fully compatible with other common multiplexing techniques such as mode-division multiplexing \cite{Oscar:20, yang2021inversedesigned, Luo2014}. Additionally, recent work has shown automated wafer-scale trimming of resonant modulators to correct for fabrication variations \cite{intel2021}, enabling future as-designed alignment of modulators and filters to the comb grid without any thermal energy consumption.  


Previous roadblocks to the use of Kerr combs for short-reach optical links have recently been overcome, with solutions for high volume manufacturability \cite{Jin2021, Liu2021, Xiang99, Ji:17, Ji:20}, high optical power per line with spectral flatness \cite{Jang2018, kim2021synchronization}, high pump-to-comb conversion efficiency \cite{Xue2015, Kim:19}, and turn-key integration in standard compact laser packages \cite{Stern2018, Shen2020}. In concert with these advancements, the work presented here represents a coming-of-age for Kerr frequency combs as optical interconnect sources, establishing them as a pragmatic contender for integration in next-generation data centres and high performance computers.

\section*{Methods}

\textbf{Active Silicon Photonic Chip.} The active silicon photonic circuit was designed using a mix of custom designed devices and verified process design kit (PDK) components provided by the foundry. Custom devices were designed using the Ansys Lumerical Finite Difference Time Domain (FDTD) Solver to extract S-parameters and create compact models which were then used to perform circuit-level simulations with the PDK devices in Ansys Lumerical INTERCONNECT. The mask layout was prepared in Cadence Virtuoso and physical verification was performed using Mentor Graphics Calibre. Fabrication was performed as part of the AIM Photonics 300 mm multi-project wafer service. The AIM process uses a 220 nm thick silicon waveguide layer with 2 $\mu$m thick buried oxide. For a single-mode strip waveguide of nominal width (480 nm), the propagation loss is approximately 2.5 dB/cm. Additionally, the AIM process and PDK use Si\textsubscript{3}N\textsubscript{4} layers for devices such as inverse-tapered edge couplers and silicon-to-Si\textsubscript{3}N\textsubscript{4} escalators, indicating that monolithic integration of the comb-generating microresonators on the same die as the active silicon devices is a future possibility with improved foundry nitride processing. More information regarding the AIM process and PDK can be found in ref. \cite{AIMPDK}.  \\

\textbf{Si\textsubscript{3}N\textsubscript{4} Chip Fabrication and Design.} Starting from a 4-inch silicon wafer, we thermally grow a 4-$\mu$m thick oxide layer as the bottom cladding. Si\textsubscript{3}N\textsubscript{4} is deposited using low-pressure chemical vapor deposition (LPCVD) in two steps and annealed at 1,200 °C in an argon atmosphere for 3 hours in between steps. After Si\textsubscript{3}N\textsubscript{4} deposition, we deposit a SiO\textsubscript{2} hard mask using plasma-enhanced chemical vapor deposition (PECVD). We pattern our devices using electron beam lithography. Ma-N 2403 resist was used to write the pattern and the nitride film was etched in an inductively coupled plasma reactive ion etcher (ICP RIE) using a combination of CHF\textsubscript{3}, N\textsubscript{2}, and O\textsubscript{2} gases. After stripping the oxide mask, we anneal the devices again to remove residual N-H bonds in the Si\textsubscript{3}N\textsubscript{4} film. We clad the devices with 500 nm of high-temperature silicon dioxide (HTO) deposited at 800 °C followed by 2 $\mu$m of SiO\textsubscript{2} using PECVD. A deep etched facet and an inverse taper are designed and used to minimize the edge coupling loss. More detailed information on the device fabrication can be found in refs. \cite{Ji:17, Ji:2021}. The coupled-microresonator system was designed to generate a 200 GHz-spaced comb and the cross-section of the Si\textsubscript{3}N\textsubscript{4} microrings was 730 x 1,000 nm. Platinum heaters were implemented over both resonators to tune an avoided mode crossing for generating a normal-GVD comb with an arbitrary pump wavelength. More information regarding the turn-key operation and noise properties of a similar comb can be found in ref. \cite{Kim:19}. \\

\textbf{Device Packaging.} The electrical and optical packaging was performed by Optelligent, LLC. The active silicon photonic chip was first die-bonded to a co-designed printed circuit board (PCB). The PCB bond pads were finished with electroless nickel electroless palladium immersion gold (ENEPIG) to enable better bond reliability with gold wirebonds. All 96 DC pads around the periphery of the photonic chip were wirebonded to the PCB and the 64 internal RF pads were left exposed for probing. An 8-channel lidless v-groove fibre array with standard cleaved SMF-28 fibres (OZ Optics) was flipped upside down to provide clearance over the chip dicing trench and was attached to the 127 $\mu$m pitch edge coupler array on the photonic chip using index matching UV-cured epoxy after active alignment. The measured optical coupling loss after packaging was approximately 2.5 dB per facet.\\

\textbf{Modulator Characterisation.} Microdisk modulators available in the AIM PDK \cite{AIMPDK} were used for the transmitter with four nominal wavelength designs (1,550 nm, 1,556.4 nm, 1,562.8 nm, and 1,569.4 nm). The modulators each have an FSR of approximately 25.6 nm. Integrated thermal heaters in each modulator were used to shift the resonances to align to the comb wavelength grid, with a measured tuning efficiency of 0.5 nm/mW. Future iterations using custom modulator designs with more nominal wavelengths and high efficiency heaters will greatly reduce the expended thermal tuning power. The measured loaded quality factor (Q) of a typical modulator was 5,200 ($f_{FWHM}$ = 39 GHz at 1,550 nm) which indicates an optical 3 dB modulation bandwidth of $f_{3 dB}^{opt} = 25$ GHz using the relation $f_{3 dB}^{opt} = \sqrt{\sqrt{2} - 1} \times f_{FWHM}$ \cite{Timurdogan2014, Osgood2002}. The electro-optic 3 dB bandwidth was measured to be $f_{3dB}^{el \mhyphen opt} = 15$ GHz with a 0 V DC bias (Fig. \ref{fig:device_characterization}c) which is combined with the calculated optical modulation bandwidth to estimate the RC bandwidth through the relation $(\frac{1}{f_{3dB}^{el \mhyphen opt}})^{2} = (\frac{1}{f_{3dB}^{opt}})^{2} + (\frac{1}{f_{3dB}^{el}})^{2}$, yielding  $f_{3dB}^{el} = 37.5$ GHz \cite{Timurdogan2014}. \\

\textbf{Interleaver Characterisation and Alignment.} The RMZI interleavers have a thermo-optic phase shifter in both the ring and bottom arm to correct for phase errors due to fabrication variations. An applied power of 25 mW to the bottom arm heater yields an FSR/2 shift of the pass- and stop-bands, corresponding to a $\pi$ phase shift. The first de-interleaver stage has a 10\% power tap in the bottom arm which is looped back to an output edge coupler to align the pass- and stop-bands. A tunable laser (Keysight 81608A), lightwave measurement system (Keysight 8164B), and four channel optical power meters (Keysight N7744A) were used to sweep the (de-)interleaver spectrum while on-chip thermo-optic heaters were tuned to align the spectrum to the comb wavelength grid. Power taps were used at the end of each bus to similarly align the second stage de-interleaver and modulator resonances. A tap before the last interleaver stage was used for final alignment to recombine the comb onto a single fibre output.\\

\textbf{Data Transmission Experiments.} The Si\textsubscript{3}N\textsubscript{4} comb chip was pumped using a tunable CW laser with 7 dBm output power (Alnair Labs TLG-200) which was amplified to $\sim$ 1 W using an erbium-doped fibre amplifier (EDFA) (Amonics AEDFA-33-B-FA). The polarisation of the pump was tuned with a polarisation controller ensure that the fundamental transverse electric (TE) mode was launched into the chip edge couplers. The edge couplers on the comb chip had a measured loss of 3 dB/facet, yielding 500 mW of on-chip optical power for comb generation. The generated comb was amplified using a broadband C- and L-band EDFA (FiberLabs AMP-FL8021-CLB-22) and was sent to the transmitter package using single-mode fibre with a polarisation controller to launch TE-polarised light into the chip. The transmitter PCB was mounted to a temperature-controlled stage (Thorlabs PTC1) to thermally stabilise the chip. DC biases for thermo-optic control of the interleavers and modulators were provided by a 96 channel source measurement unit (Qontrol) connected to the PCB. The bit pattern was generated by a bit-error rate tester (BERT) (Anritsu MP1900A) and was sent to the modulators using RF multi-contact wedge probes (Cascade Infinity). All data transmission experiments used a pseudo-random bit sequence of length $2^{31} - 1$ (PRBS31) to provide the most rigorous test pattern and eliminate the possibility of pattern-dependent behaviour. After modulation, the light was coupled off the chip into single-mode fibre and amplified using a broadband C- and L-band EDFA (FiberLabs AMP-FL8021-CLB-22). A tunable bandpass filter (Finisar Waveshaper) was then used to select single modulated comb lines with a variable optical attenuator (Thorlabs EVOA1550A) to control the optical power to the receiver. The modulated comb lines were then converted back into the electrical domain using a photodiode and trans-impedance amplifier (Thorlabs RXM40AF). The received signal was sent to a real-time oscilloscope (Keysight Infiniium Z-Series) for eye characterisation or back to the BERT for bit-error rate evaluation.

\begin{acknowledgments}
This work was supported in part by the U.S. Advanced Research Projects Agency--Energy under ENLITENED Grant DE-AR000843 and in part by the U.S. Defense Advanced Research Projects Agency under PIPES Grant HR00111920014. This work was performed in part at the Cornell NanoScale Facility, a member of the National Nanotechnology Coordinated Infrastructure (NNCI), which is supported by the National Science Foundation (Grant NNCI-2025233). The authors thank Paul Gaudette and David C. Scott (Optelligent, LLC) for device packaging, AIM Photonics for chip fabrication, Michael L. Fanto for assistance with chip imaging, and Analog Photonics for PDK support. Additionally, the authors acknowledge fruitful conversations with Xiang Meng, Nathan C. Abrams, Hao Yang, and Yu-Han Hung.
\end{acknowledgments}

\section*{Author Contributions} 
A.R. and Q.C. conceived the link architecture and performed initial calculations. A.R. performed the circuit simulations, designed the floorplan, and drew the top level mask layout. V.G. designed the PCB and conducted the DC electro-optic characterisation. B.Y.K. and Y.O. designed, simulated, and characterised the Kerr comb microresonators and X.J. fabricated the Si\textsubscript{3}N\textsubscript{4} chip. A.N. led the high-speed testing and data transmission experiments with assistance from A.R., S.D., and V.G. All authors helped analyse the data and A.R. prepared the manuscript with input from all authors. M.L., A.L.G., and K.B. supervised the project. \\


\section*{Competing Interests}
The authors declare no competing interests. 

\section*{Data Availability}

The data that support the plots within this paper and other findings of this study are
available from the corresponding author upon reasonable request.

\section*{Correspondence}

Correspondence and requests for materials should be addressed to K.B. (email: bergman@ee.columbia.edu).

\bibliographystyle{naturemag}
\bibliography{main.bbl}

\clearpage


\section*{Supplementary material}

# Supplementary Information: Integrated Kerr frequency comb-driven silicon photonic transmitter

Anthony Rizzo[1,†], Asher Novick[1,†], Vignesh Gopal[1,†], Bok Young Kim[2], Xingchen Ji[1], Stuart Daudlin[1], Yoshitomo Okawachi[2], Qixiang Cheng[1,3], Michal Lipson[1], Alexander L. Gaeta[1,2], and Keren Bergman[1,*]

[1]*Department of Electrical Engineering, Columbia University, New York, NY 10027*
[2]*Department of Applied Physics and Applied Mathematics, Columbia University, New York, NY 10027*
[3]*Current Address: Engineering Department, University of Cambridge, Cambridge CB3 0FA, UK*
[†]*These authors contributed equally to this work*
(Dated: September 7, 2021)

## 1. SUPPLEMENTARY NOTE: OPTICAL LINK BUDGET

To demonstrate the scalability of our proposed silicon photonic link architecture to terabit/s/fibre capacities, we provide a comprehensive analytical modelling framework for the optical link budget. We consider a proposed architecture of 64 wavelength channels at 16 Gb/s/$\lambda$, yielding an aggregate single fibre bandwidth of 1.024 Tb/s. We select a frequency comb channel spacing of 150 GHz and use 2 interleaver stages, resulting in four 16 channel groups with 600 GHz channel spacing per group at the resonant modulator and filter arrays. To determine the optical link budget we consider the insertion loss (IL) of the edge couplers, interleavers, modulator arrays, and filter arrays in addition to the analytical power penalty models associated with these devices.

Previous state-of-the-art demonstrations for silicon photonic edge couplers have shown a broadband coupling IL ≤ 1 dB [1]. Adhering to works which use a comparable material stack and fabrication process to our devices, we assume 1 dB IL per facet. Our proposed link architecture uses ring-assisted Mach Zehnder interferometers (RMZIs) as tunable flat-top even-odd channel interleavers and de-interleavers. The bandwidth of these devices is typically limited by their least broadband component. To model their bandwidth dependent loss, we measured RMZI interleaver transmission spectra on test chips and fit a polynomial curve to the peaks to model the IL (Fig. S1).

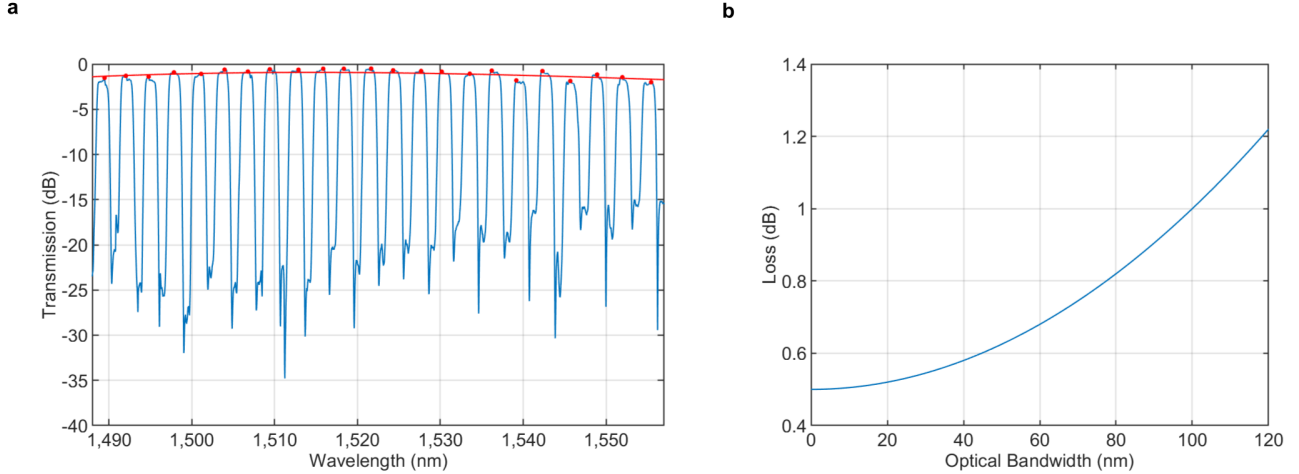

FIG. S1: **Modelling the broadband IL of RMZI-based interleavers. a**, Measured RMZI tranmission spectrum (blue) with polynomial loss envelope fit to peaks (red). **b**, IL as a function of optical bandwidth using the fitted polynomial from **a**.

The 64 channel 150 GHz spaced comb utilizes 9.6 THz of optical bandwidth ($\Delta f$). We choose our center wavelength to be $\lambda_c = 1{,}510$ nm to maximize overlap with regions of high photodiode responsivity and edge coupler efficiency, which tend to have higher performance at shorter wavelengths. We calculate our utilized optical bandwidth in terms of wavelength by:

$$\Delta\lambda = \frac{\lambda_c^2}{c}\Delta f \tag{1}$$

where $\Delta\lambda$ is the optical bandwidth in metres, $\lambda_c$ is the centre wavelength in metres, $c$ is the speed of light in vacuum, and $\Delta f$ is the optical bandwidth in hertz. Thus, $\Delta\lambda = 73$ nm, which corresponds to $\approx 0.7$ dB loss per interleaver stage from the polynomial fit show in Fig. S1b.

For our 16 channel resonator array we consider the on-resonance insertion loss of the target resonator aligned to the carrier wavelength of a channel as well as the off-resonance loss contributed by the 15 additional resonators. For the former, on-resonance insertion losses as low as 1 dB and and 0.2 dB have been demonstrated in past works [2, 3] for microdisk modulators and microring resonators respectively. For the latter, we have measured the off-resonance insertion loss as nearly negligible and conservatively assume as high as 0.05 dB per off-resonance channel.

Additionally, we assume $\approx 1$ dB/cm propagation loss due to the bus waveguides throughout the silicon photonic link [3]. The primary contributors to propagation length are the resonator arrays, which require a pair of RF contact pads per resonator in close spatial proximity to each resonator. For 16 channel arrays with the RF contact pads at 50 $\mu$m pitch, there is $\approx 1.6$ mm routing length per array. Therefore, we conservatively consider our total waveguide routing length to be 3.5 mm (slightly greater than the sum of a pair of modulating and filtering resonator arrays).

The first power penalty is the non-return-to-zero on-off keying (NRZ-OOK) penalty. There are two components to calculate, both entirely determined by the extinction ratio (ER), $r_{ER}$, of the signal, which is defined as:

$$r_{ER} = \frac{P_1}{P_0} \qquad (2)$$

where $P_1$ and $P_0$ are the mean optical powers of a '1' and '0', respectively. For an ideal OOK signal we have $r_{ER} = \infty$. As the ER degrades to lesser values, we begin to see an impact on the measured signal quality at the receiver. We model this logarithmic power penalty, $PP_{ER/OOK}$, as [4]:

$$PP_{ER/OOK} = PP_{ER} + PP_{OOK} = -10\log_{10}\left(\frac{r_{ER}-1}{r_{ER}+1}\right) - 10\log_{10}\left(\frac{r_{ER}+1}{2r_{ER}}\right) \qquad (3)$$

We select microdisk modulators with ER = 8 dB and an on-resonance insertion loss of 1 dB, as has been demonstrated in past work [2]. Fig. S2 visualizes equation (3). For the assumed extinction ratio of 8 dB, the associated power penalty is 3.75 dB.

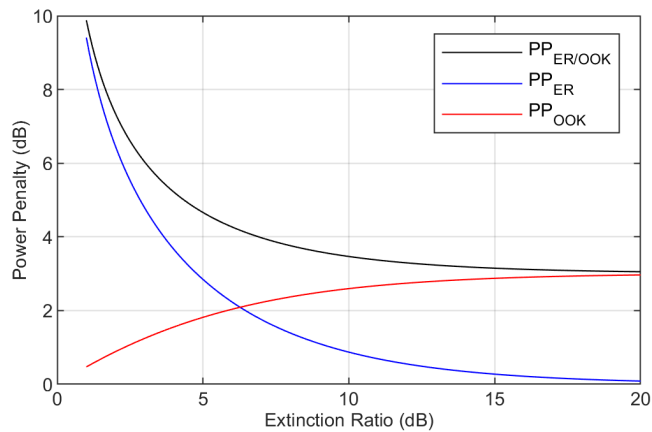

FIG. S2: **Modulator power penalty as a function of extinction ratio.** Plot of equation (3) on logarithmic axes. As ER $\to \infty$, $PP_{ER/OOK} \to 3$ dB. The power penalty at ER = 8 dB is 3.75 dB.

The second power penalty we calculate is the system crosstalk. We consider microdisk modulators and microring filters with $R \approx 4.5$ $\mu$m which results in a free spectral range (FSR) of $\approx 3.2$ THz. As calculated earlier, this results in utilizing 3 FSRs worth of optical bandwidth. Supplementary Note 4 describes in detail how the link can operate in this multi-FSR regime. $\Delta\lambda_{aggressor}$ is defined as the spacing between any given channel and the nearest adjacent resonance corresponding to a different channel. Applying equation (7) (Supplementary Note 4) to this FSR with 16 channels per group yields:

$$\Delta\lambda_{aggressor} \approx \frac{3.2 \text{ THz}}{16} = 200 \text{ GHz}. \qquad (4)$$

Previous works have shown the power penalty of resonator-based WDM links is negligible when $\Delta\lambda_{aggressor} \geq 200$ GHz [5]. Thus, the proposed multi-FSR regime link is minimally affected by inter-channel crosstalk at the modulators and receivers.

Finally, we consider the impact of the filtering done by the add-drop microring at the receiver. It has been shown in previous works that a microring resonator filter can both equalize an NRZ-OOK signal and filter out undesired optical noise [6]. The combined effect has been measured as improved detection sensitivity by as much as 4.5 dB under the correct set of conditions. This improved detection sensitivity corresponds to a *negative* power penalty. Pending further work on modelling this effect, we conservatively assume a signal improvement of 1 dB.

Table S1 tabulates the total optical link budget, including the impact of device insertion loss and abstract power penalties.

| **Transmitter** | | | |
|---|---|---|---|
| Coupler IL [dB] | Interleaver IL [dB] | Disk Modulator Array IL [dB] | Signal Penalties [dB] |
| 2 | 2.80 | 1.75 | 3.75 |
| **Receiver** | | | |
| Coupler IL [dB] | Interleaver IL [dB] | Ring Filter Array IL [dB] | Signal Penalties [dB] |
| 1 | 1.4 | 0.95 | -1 |
| Propagation Loss [dB] | Tx Total [dB] | Rx Total [dB] | **Total [dB]** |
| 0.35 | 10.3 | 2.35 | **13** |

TABLE S1: **Sum of total insertion losses and power penalties in the proposed 1.024 Tb/s link.**

## 2. SUPPLEMENTARY NOTE: ESTIMATED ENERGY CONSUMPTION

To determine the estimated energy consumption we first refer to our optical link budget in Supplementary Note 1. Using state-of-the-art receiver designs we propose a receiver sensitivity threshold of $\leq -19$ dBm at a channel data rate of 16 Gb/s [7]. Adding the estimated optical link penalty of a 16 Gb/s by 64 channel link to this results in a minimum optical power per line for our comb of -6 dBm.

Coupled-ring normal group-velocity dispersion (GVD) regime Kerr frequency comb sources have demonstrated up to 41% pump-to-comb conversion efficiency [8]. Combined with the pump laser wall plug efficiency of 35% for a high-powered distributed feedback (DFB) pump [9], the integrated DWDM source has an overall efficiency of 14.35%.

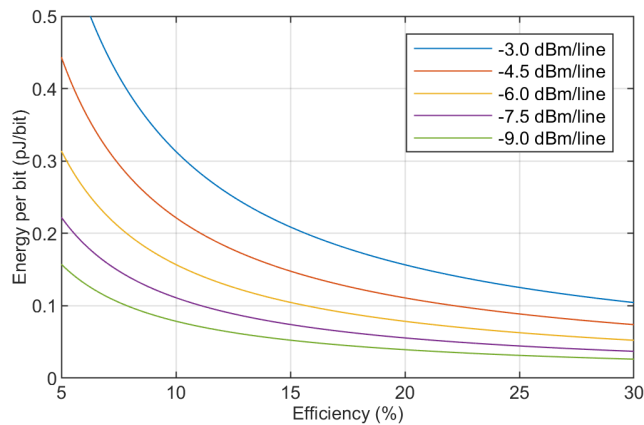

FIG. S3: **Comb energy per bit.** For a 1.024 Tbps link architecture comprised of 64 comb lines, each modulated at 16 Gbps, -6 dBm per line at 14.35% efficiency corresponds to 109 fJ/bit.

For modulation with an NRZ-OOK PRBS signal, 0-to-0, 0-to-1, 1-to-0, and 1-to-1 transitions are all equally probable, resulting in a modulator energy consumption per bit of [10]:

$$E_{bit} = \frac{1}{4}CV_{drive}^2 \tag{5}$$

where $E_{bit}$ is the consumed energy for transmitting one bit, $C$ is the lumped device capacitance, and $V_{drive}$ is the peak-to-peak driving voltage applied to the modulator terminals.

Previous microdisk modulator demonstrations have shown $C \leq 50$ fF while being driven with $V_{drive} \leq 1$ V [2]. Applying these numbers to equation (5) results in 12.5 fJ/bit energy consumption from the microdisk modulators. In addition to this, the aforementioned state-of-the-art receiver design has been demonstrated as consuming roughly 180 fJ/bit [7].

Nearly every resonator in the architecture requires thermal tuning to align their resonances, either to the comb pump or to a wavelength channel. The coupled-ring normal GVD regime Kerr comb has been shown to require $\approx$ 100 mW of thermal tuning energy, corresponding to 98 fJ/bit for our 1.024 Tb/s capacity link. For the microdisk modulators and microring filters, through using a substrate undercut to thermally isolate devices, thermal tuning efficiency as low as 2.4 mW/FSR ($P_\pi = 1.2$ mW) can be achieved [11]. It has been shown that vertical junction microdisk modulator resonance variation across a 300 mm wafer can suffer from a $\sigma(\Delta\lambda) = 2.14$ nm [3]. Applying the $P_\pi$ from the thermal undercut and assuming we may require tuning the 4.5 $\mu m$ radius disks over $3\sigma$ (covering 99.7% of the distribution), we have 0.6 mW required thermal tuning power per resonator. At 16 Gpbs, this corresponds to 37.5 fJ/bit. Applying the same reasoning to the microring resonators and adding the comb tuning results in a cumulative thermal tuning of 173 fJ/bit. Recently, promising results for wafer-scale post-fabrication trimming of resonant modulators have been shown for near-perfect correction of fabrication variations in the resonant wavelengths [12]. Through applying these techniques, future systems could dramatically reduce the thermal tuning budget to near-zero for fabrication error correction. Small amounts of thermal tuning will still be necessary to correct for chip temperature fluctuations; however, promising results using CMOS-compatible titanium oxide cladding to neutralize the strong thermo-optic effect in silicon [13] provide a path to entirely eliminating thermal tuning.

Table S2 tabulates the total estimated energy consumption for the proposed link.

| Laser [pJ/bit] | Modulator [pJ/bit] | Receiver [pJ/bit] | Thermal Tuning [pJ/bit] | **Total [pJ/bit]** |
|---:|---:|---:|---:|---:|
| 0.109 | 0.0125 | 0.180 | 0.173 | **0.4745** |

TABLE S2: **Sum of total energy consumption for 1.024 Tb/s link, comprised of 64 comb lines spaced by 150 GHz and modulated at 16 Gb/s each.**

### 3. SUPPLEMENTARY NOTE: KERR COMB STABILITY AND NOISE

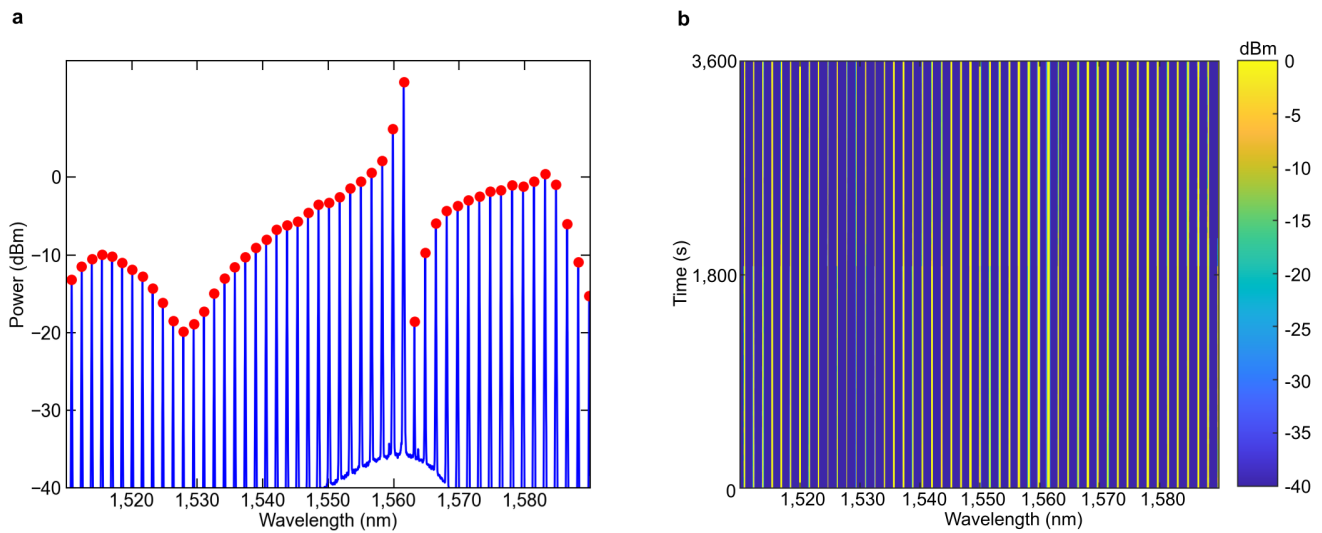

FIG. S4: **Comb stability over time. a**, Measured comb spectrum in the low noise state with peaks annotated. **b**, Measured power per line of comb in **a**, measured every 30 seconds for 1 hour with an optical power scale saturation of 0 dBm for visual clarity. No closed loop thermal control is used, demonstrating long term line stability despite environmental fluctuations.

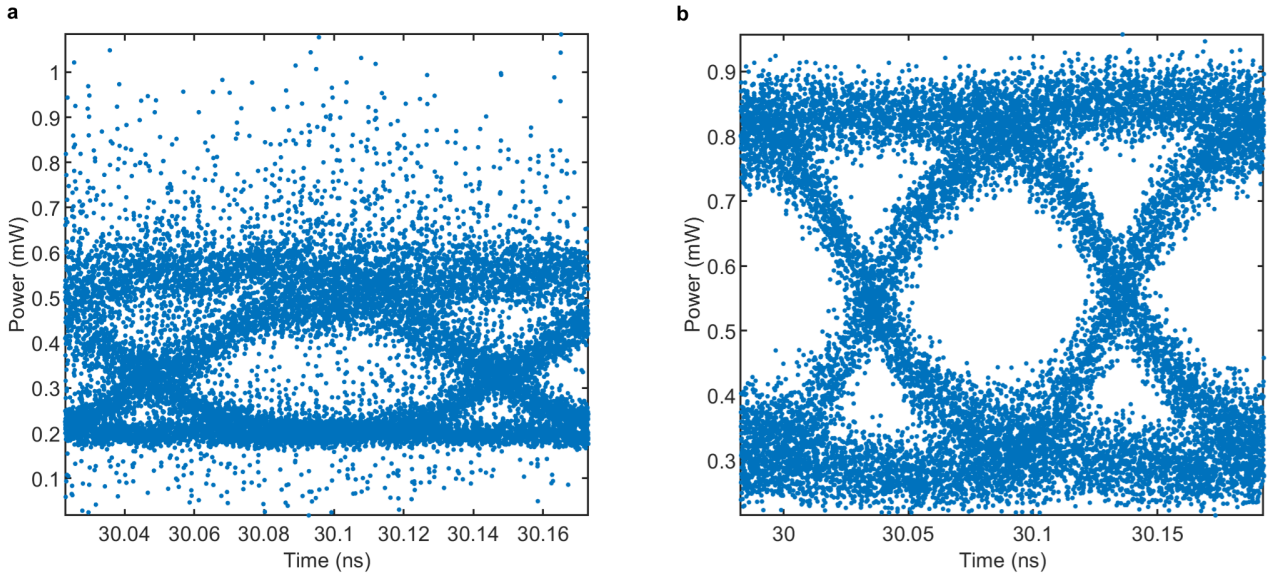

FIG. S5: **Impact of comb state on signal quality. a**, Measured eye diagram of a comb line ($\lambda = 1{,}559.8$ nm) at 10 Gb/s when the comb is in the high-noise state, with a quality factor (Q) of $\approx 2.5$ and an estimated bit error rate (BER) of $6 \times 10^{-3}$. **b**, Measured eye diagram of the same comb line at 10 Gb/s when the comb is in the low-noise state, with a Q of $\approx 5.8$ and an estimated BER of $4 \times 10^{-9}$.

## 4. SUPPLEMENTARY NOTE: SCALING BEYOND THE SINGLE-FSR REGIME

Single bus cascaded resonator links are fundamentally restricted by the limited FSR of the microresonators, preventing further channels from being added due to the 'alias' resonances of existing channels (Fig. S6a). While the FSRs of these resonators can be increased by making them smaller, there are physical limitations to how small their radii can be before incurring large losses due to bend radiation [2]. Furthermore, although more channels can be added by decreasing the channel spacing, previous works have shown that channel spacings below 100 GHz lead to severe crosstalk penalties [5, 14]. Due to the broad spectral bandwidth spanned by microresonator Kerr frequency combs, alternate channel arrangement schemes beyond the single-FSR regime are required for scaling to large channel counts.

Our proposed multi-FSR architecture employs RMZI-based interleavers due to ease of device implementation in high volume fabrication processes and device footprint for achieving large bandwidth densities. These devices split the comb into 'even' and 'odd' groups at each stage with flat-top pass- and stop-bands, allowing for misalignment tolerance with the comb. Since this splitting does not reduce the total optical bandwidth spanned on each bus which is much larger than the resonator FSR, the resonator arrangement must be chosen such that the 'alias' resonances do not lead to crosstalk with other channels (Fig. S6). Since each stage of de-interleaving doubles the comb spacing and halves the channel count, it is possible to construct these multi-FSR channel arrangements for large channel counts in a closed-form manner when evaluated holistically with the rest of the link architecture.

In the absence of dispersion, the characteristic relations for the multi-FSR regime are given by:

$$\Delta\lambda = (1 + \frac{1}{N})(m * FSR) \tag{6}$$

$$\Delta\lambda_{aggressor} = \frac{FSR}{N} \tag{7}$$

where $\Delta\lambda$ is the channel spacing on a single bus, $N$ is the number of wavelength channels on a single bus, $m$ is an integer, $FSR$ is the free spectral range of the resonators, and $\Delta\lambda_{aggressor}$ is the spacing between the target resonance and the nearest 'alias' resonance. With equations (6) and (7), one can set the minimum guard bands through $\Delta\lambda_{aggressor}$ and evaluate combinations of $FSR$, $N$, and $\Delta\lambda$ which satisfy the constraints. Of these combinations, $N$ will be determined by the total number of comb lines and number of interleaver stages and $\Delta\lambda$ will be determined by the comb spacing and number of interleaver stages. Typically, multiple satisfactory link architectures exist,

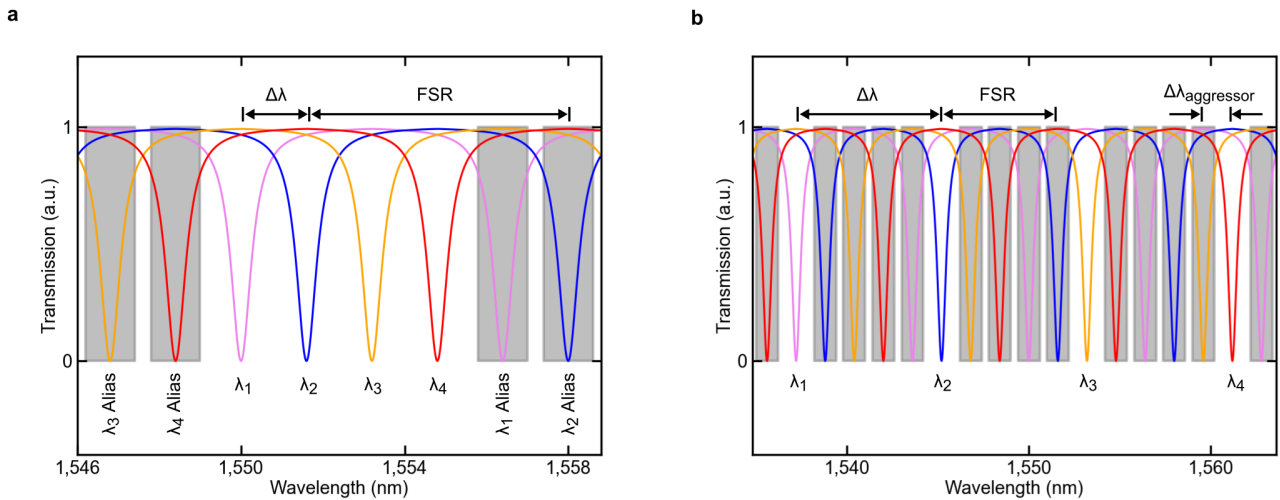

FIG. S6: **Generic channel arrangements for the single- and multi-FSR regimes. a**, Standard single bus channel arrangement where all target channels are within a single FSR of the resonators. **b**, Multi-FSR channel arrangement scheme that uses a Vernier effect-like configuration such that the 'alias' resonances of the modulators are slightly misaligned from the other target channels, largely avoiding crosstalk effects. In the single-FSR regime, the channel spacing $\Delta\lambda$ determines the crosstalk characteristics whereas for the multi-FSR regime, the crosstalk is determined by $\Delta\lambda_{\text{aggressor}}$.

and additional pragmatic constraints must be considered such as physically achievable comb spacings, the maximum number of channels provided by the comb, and the total insertion loss as a function of interleaver stages. Furthermore, while these relations algorithmically yield valid link architectures, additional valid configurations exist without a closed-form solution; these are necessary to examine if there is a restriction on the degrees of freedom, such as if the resonator FSR is fixed. For example, Fig. S7 shows the valid four bus architecture detailed in Supplementary Note 1 with 16 channels per bus and 150 GHz comb spacing using the fixed AIM PDK microresonator FSRs of 25.6 nm. In this case, $\Delta\lambda_{aggressor} = 200$ GHz, $N = 16$, and $\Delta\lambda = 4.8$ nm ($\approx 600$ GHz at 1550 nm), which do not satisfy the relations yet result in a usable link architecture.

Limitations of this regime are the stochastic variations in FSR due to fabrication variations as well as the impact of dispersion. To first order, the impact of dispersion on FSR is given by [15]:

$$\text{FSR} = \frac{\lambda^2}{n_g(2\pi R)} \tag{8}$$

where $\lambda$ is wavelength, $R$ is the radius of the disk, and $n_g$ is the group index of the fundamental transverse electric (TE) whispering gallery mode. Since silicon is a highly dispersive material at telecommunication wavelengths, $n_g$ is non-constant over large bandwidths and thus the resonator FSR is wavelength-dependent (Fig. S8). While the closed-form solution given by equations (6) and (7) and Fig. S6b are useful visualization tools, realistic link configurations for fabrication must be evaluated numerically to include the effects of dispersion. In addition to dispersion considerations, since the FSR depends on $n_g$ which is difficult to tune post-fabrication, it is necessary to quantify the effect of fabrication variations for the multi-FSR regime.

Using experimental fabrication variation probability distributions provided by the foundry [3], 10,000 Monte Carlo simulations were run for both the modulator and interleaver to determine the impact of fabrication variations on the device FSR (Fig. S9). The captured statistics were comprehensive to all sources of variation including silicon width, height, partial etch height, and doping fluctuations. The resulting FSR probability distributions show promise for high yield in terms of channel arrangement variation, with a standard deviation ($\sigma$) mismatch of 0.32 nm/FSR for the modulators and 0.96 GHz/FSR for the interleavers. These numbers can additionally be improved through careful engineering of the resonators and interferometers with compensating width sections [16]. For the simulated resonators, assuming a $\Delta\lambda_{\text{aggressor}}$ of 300 GHz ($\approx 2.4$ nm at $\lambda = 1550$ nm) and 50 GHz guard bands, the multi-FSR regime can accommodate a 3-FSR $\sigma$ walk-off (3 FSRs × 0.32 nm/FSR = 0.96 nm = 120 GHz) without infringing on the worst-case guard bands. Furthermore, resonant and interferometric silicon photonic devices in close spatial proximity have been shown to have high correlation in their variations [17], which further reduces the potential for channel spacing mismatch within a single chip.

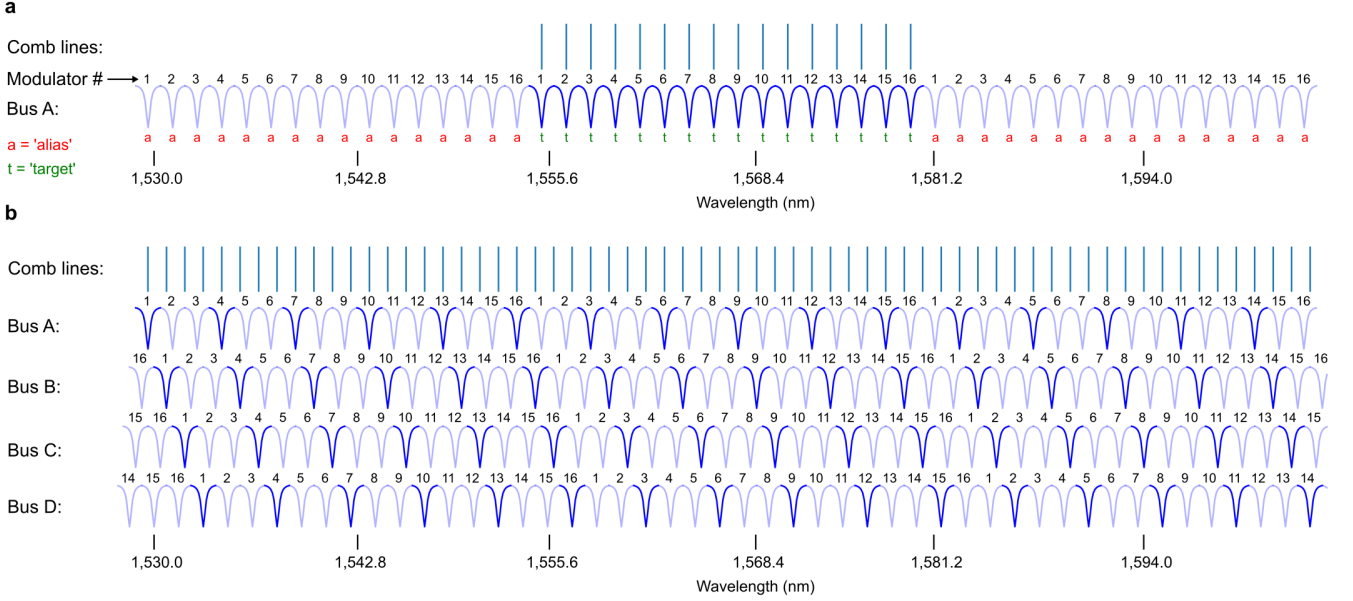

FIG. S7: **Example channel arrangement for 64 wavelengths across 4 buses.** **a**, Schematic of a typical single-FSR cascaded resonator link for 16 channels with 'alias' and 'target' channels labelled. The resonator FSR is 25.6 nm and the comb spacing of 150 GHz is equal to the resonance channel spacing. **b**, Enumerated example of a multi-FSR regime channel arrangement for 64 channels split into four bus waveguides. Again, the resonator FSR is 25.6 nm and the comb spacing is 150 GHz ($\approx$ 1.2 nm at 1550 nm) but $\Delta\lambda_{aggressor} = 1.6$ nm, yielding reduced crosstalk compared to the single bus case. The resonance linewidth is exaggerated compared to its actual physical value for visual clarity.

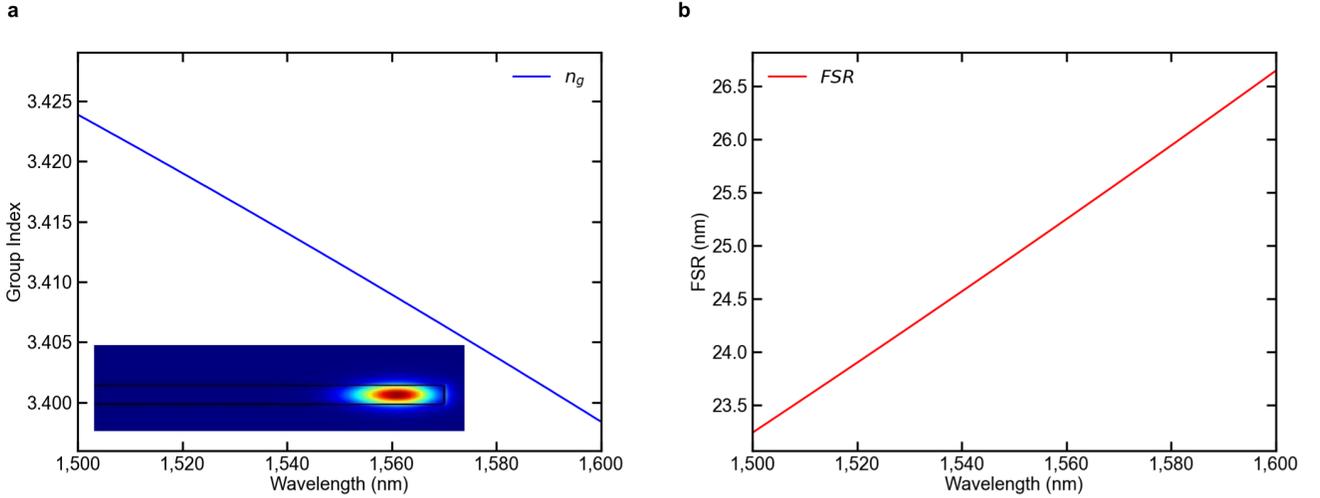

FIG. S8: **Quantification of the impact of first-order dispersion on the microdisk modulator FSR.** **a**, Simulated group index $n_g$ for a 4.5 $\mu$m radius microdisk over 100 nm spectral bandwidth. Inset: cross-sectional mode profile of the fundamental whispering gallery mode from finite difference eigenmode simulations. **b**, Corresponding FSR as a function of wavelength using the simulated $n_g$ values and equation (8).

Alternate channel splitting schemes for subdividing the comb such as using chip-based dichroic filters are possible [18], but the slow roll-off in demonstrated integrated devices would result in unusable comb lines at the transition edge. Additional demonstrations have shown schemes for FSR-free resonant modulators [19, 20] and FSR-free resonant filters [21] through the use of more complicated coupling structures blended with subwavelength gratings to create additional spectral dependence. Furthermore, through combining a small-radius microdisk with a mode-selective bent

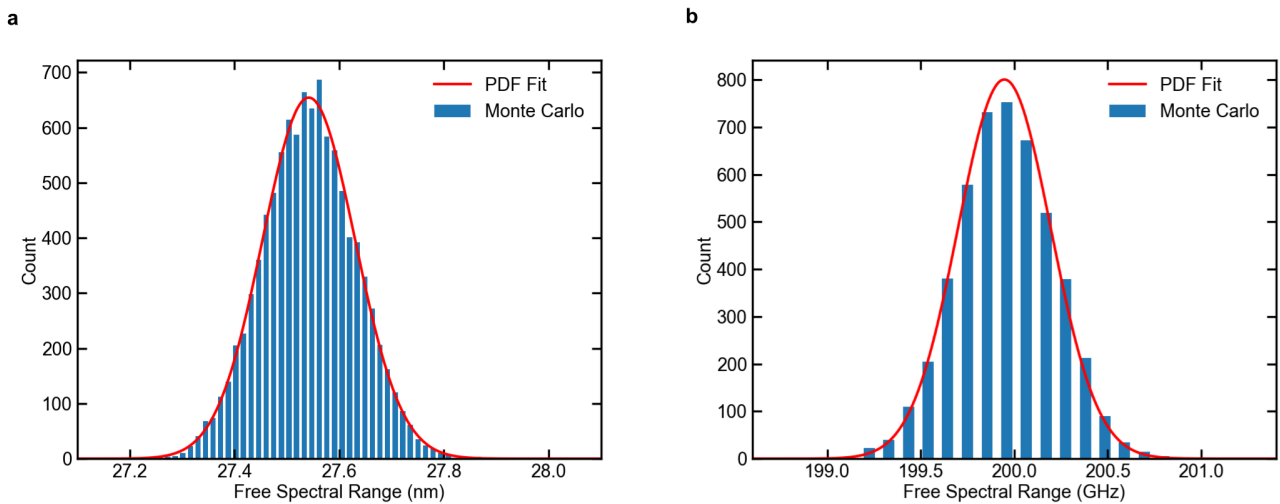

FIG. S9: **Analysis of impact of fabrication variations on critical device parameters for the multi-FSR regime.** Results from 10,000 Monte Carlo simulations in Lumerical INTERCONNECT for fabrication variations in the (a) microdisk modulator FSR and (b) interleaver FSR. All simulations use statistical data provided by the foundry for wafer-scale measured experimental data. PDF: probability density function.

directional coupler, a microdisk modulator with an FSR of 86 nm was demonstrated [22]. If the insertion losses of such devices can be made sufficiently low, it may be possible to extend the standard single bus configuration to extremely large channel counts which are not restricted by the modulator/filter FSRs.



\end{document}